\documentstyle[preprint,prb,aps,epsf]{revtex}
\def\inseps#1#2{\def\epsfsize##1##2{#2##1} \centerline{\epsfbox{#1}}}
\begin{document}
\draft
\input{psfig}
\title{Lattice Boltzmann simulations of lamellar and droplet phases}
\author{G. Gonnella$^1$, E. Orlandini$^2$,
and J.M. Yeomans$^3$}
\address{$^1$Dipartimento di Fisica dell'Universita' di Bari and
Istituto Nazionale di Fisica Nucleare, Sezione di Bari, via Amendola
173, 70126 Bari, Italy.\\
$^2$Service de Physique Theorique, Centre d'etudes
de Saclay, F-91191 Gif-sur-Yvette, France.\\
$^3$Theoretical Physics, Oxford University,
1 Keble Rd., Oxford OX1 3NP, UK.
}
\date{\today}
\maketitle
\begin{abstract}

Lattice Boltzmann simulations are used to investigate spinodal
decomposition in a two-dimensional binary fluid with equilibrium
lamellar and droplet phases. We emphasise the importance of
hydrodynamic flow to the phase separation kinetics.
For mixtures slightly asymmetric in composition the fluid phase
separates into bulk and lamellar phases with the lamellae forming
distinctive spiral structures to minimise their elastic energy.

\end{abstract}


\newpage

\section{Introduction}

Complex fluids such as microemulsions, foams, and colloidal suspensions
provide a wealth of interesting physical phenomena. Their use is
ubiquitous in the processing, energy and chemical
industries. Therefore it is important to explore ways in which their
properties can be modelled numerically to test physical theories,
predict behaviour and provide feedback to industrial planning.

The modelling of complex fluids is not an easy task as both the
rheology and the complicated phase behaviour of the fluid must be
incorporated. Molecular dynamics simulations, whilst providing an
accurate picture of the microscopic physics, are usually too
computer-intensive to address hydrodynamic time scales. In
computational fluid dynamic solutions of the Navier-Stokes equations
the specific behaviour of a given fluid can only be input via ad hoc
constitutive relations.

Recently new methods of simulating fluid flow have become
available. These include lattice-gas cellular
automata\cite{EC97}, 
dissipative
particle dynamics\cite{HK92}, 
and lattice Boltzmann simulations\cite{BS92}. The aim is to
reproduce the physics of fluid flow, primarily mass and momentum conservation,
while including the important features of the underlying microscopic
physics. These approaches have been successfully applied to several
complex fluids including polymer solutions\cite{SH95}, 
particulate suspensions\cite{BC96}
and microemulsions\cite{EC97}. However 
application of the techniques to model
complex fluids is still in its infancy and validation of, and comparison
between, the different methods is still needed.

In this paper we concentrate on lattice Boltzmann simulations. A
fluid is modelled on a mesoscopic length scale by means of
distribution functions which evolve according to a
discretised version of a simplified Boltzmann equation. The correct
equilibrium behaviour is imposed by inputting the correct
thermodynamics such that the system evolves to the minimum of a chosen
input free energy\cite{SO95,OS95}.

We use the lattice Boltzmann approach
to explore the kinetics of phase separation of a model, two-dimensional, 
fluid with lamellar and droplet
equilibrium phases. Our main conclusions are:
\begin{enumerate}
\item When the system is quenched to the lamellar phase a glassy 
metastable state of tangled lamellae forms.  However, hydrodynamic 
flow alleviates the frustration and allows the system to attain the 
striped ground state.
\item As the concentration of the minority phase is increased 
the system prefers to phase separate into a lamellar 
and a one-component phase. The lamellae minimise 
their free energy by forming a spiral pattern around the one-component
hole.  Again this state can only be reached by the fluid if
hydrodynamic  flow is allowed in the system.
\item When one component of the binary fluid predominates 
droplets of the minority fluid form.  Their size is determined by 
the balance of the surface tension terms in the free energy.
Phase separation in this system is unaffected by hydrodynamics.

\end{enumerate}

Sections 2 and 3 of the paper are devoted to a description of the
model and the relevant thermodynamics and to a summary of the
numerical approach respectively. 
Section 4 summarises phase separation in a 50:50
fluid mixture emphasising the r\^{o}le of hydrodynamics. The 60:40
mixture which phase separates into a lamellar and one-component state
is considered in Section 5. Section 6 treats the 90:10 composition
ratio where droplets of the minority phase are stable. Section 7
summarises the results, putting them in the context of previous work,
and suggests directions for future research.

\section{Model}

We consider a two-dimensional binary fluid with components $A$, $B$ 
of number density $n_A$, $n_B$ respectively.  This can be modelled by 
the Landau free energy\cite{B75}
\begin{equation}
\Psi=\int d\vec{r}\left\{ \frac{a}{2} (\varphi)^2
+\frac{b}{4} (\varphi)^4
+\frac{\kappa}{2}(\nabla \varphi)^2 
+\frac{\zeta}{2}(\nabla^2\varphi)^2\right\}
\label{eqn1}
\end{equation}
where $\varphi=n_A-n_B$ is the order parameter of the system and we 
assume that the total density $n =n_A + n_B$ is constant.  The free 
energy (\ref{eqn1}) corresponds to a disordered state at high 
temperatures ($a>0$) and an ordered state at low temperatures 
($a<0$).  

$\kappa$ is related to the surface tension.  For $\kappa$ 
sufficiently large the ordered phase consists of two coexisting 
bulk phases with $n_A-n_B=\pm \varphi_0$, the volume of each
determined by the initial ratio of $n_A$ and $n_B$.  
As $\kappa$ 
is decreased the formation of interfaces becomes favourable but the
fluid remains stable because of the positive curvature energy related
to $\zeta$ in (\ref{eqn1}).  For $n_A
\sim n_B$ the 
ordered state is then a striped or lamellar phase
with the width of the lamellae being determined by the
interface--interface interaction.
For $n_A \ll n_B$ this is replaced by a phase of circular droplets of 
$A$ in $B$ with radius
determined by the balance between the negative surface tension and the
positive curvature free energies of the interfaces.

Thermodynamic properties of the binary fluid follow directly from 
the free energy (\ref{eqn1}).  In particular we shall need the 
chemical potential $\Delta \mu$ 
which couples to the density difference $\varphi$
\begin{equation}
\Delta \mu =\frac{\delta \psi}{\delta \varphi}= a \varphi+b \varphi^3
-\kappa \nabla^2 \varphi  +\zeta (\nabla^2)^2 \varphi.
\label{eqn3}
\end{equation}
Obtaining the pressure tensor  is slightly more complicated\cite{RW82}.
The pressure parallel to the interface follows from (\ref{eqn1})
\begin{equation}
p_L=\frac{a}{2} \varphi^2 +\frac {3b}{4} \varphi^4
-\kappa \varphi (\nabla^2 \varphi)
-\frac{\kappa}{2} (\nabla \varphi)^2
+\zeta \varphi (\nabla^2)^2
\varphi-\frac{\zeta}{2}(\nabla ^2 \varphi)^2.
\end{equation}
However off-diagonal terms must be added to ensure that the 
pressure tensor obeys the equilibrium condition
\begin{equation}
\partial_{\alpha} P_{\alpha\beta}=0.
\end{equation}
Considering a linear combination of all symmetric tensors 
having two or four gradient operators shows that a suitable choice 
is 
\begin{eqnarray}
P_{\alpha\beta}&=&
\{p_L +\zeta(\nabla^2 \varphi)^2 
+\zeta \partial_\gamma \varphi \partial_\gamma (\nabla^2 \varphi)\}
\delta_{\alpha\beta} \nonumber \\ 
&&+\kappa\partial_\alpha \varphi \partial_\beta \varphi
-\zeta\left\{ \partial_\alpha \varphi \partial_\beta(\nabla^2\varphi)
+\partial_\beta \varphi \partial_\alpha(\nabla^2\varphi)\right\}.
\label{eqn5}
\end{eqnarray}

\section{Lattice Boltzmann simulations}

The aim is to simulate a fluid with equilibrium properties described
by the free energy (\ref{eqn1}) which obeys the Navier-Stokes and
convection-diffusion equations. To this end we use a lattice Boltzmann
technique\cite{OS95}.

We consider a square lattice and define two sets of distribution
functions $\{f_i\}$ and $\{g_i\}$ on each lattice site $\vec{x}$. Each
$f_i,g_i$ is associated with a lattice vector $\vec{e}_i$.
The results presented in this paper are for a 9-velocity model on a
square lattice with $e_i/c=(\pm1,0),(0,\pm1),(\pm 1/\sqrt 2,\pm 1/\sqrt
2), (0,0)$.

Physical variables are related to the distribution 
functions through
\begin{eqnarray}
n=\sum_i f_i,&\ \ \ \ \ \ &n u_{\alpha}=\sum_i f_i e_{i \alpha},
\label{5.3}\\
&\varphi = \sum_i g_i&
\label{5.4}
\end{eqnarray}
where $\vec{u}$ is the mean fluid velocity.

The distribution functions undergo a collision step followed by a
streaming step according to the evolution equations 
\begin{equation}
f_{i}(\vec{x}+\vec{e}_{i}\Delta t,t+\Delta t) - f_{i}(\vec{x},t)=
-\frac{1}{\tau_1} ( f_{i}- f_{i}^{0}),
\label{eqn6}
\end{equation}
\begin{equation}
g_{i}(\vec{x}+\vec{e}_{i}\Delta t,t+\Delta t) - g_{i}(\vec{x},t)=
-\frac{1}{\tau_2} ( g_{i}- g_{i}^0)
\label{eqn7}
\end{equation}
where $\Delta t$ is the time step, $\tau_1$ and $\tau_2$ are
relaxation parameters, and $f_i^0$ and $g_i^0$ are equilibrium
distribution functions the choice of which determines the physics
inherent in the simulation. Equations (\ref{eqn6}) and (\ref{eqn7})
are discrete Boltzmann equations with a BGK collision term\cite{BG54}.

Following the standard lattice Boltzmann prescription we assume that
$f_i^0$ and $g_i^0$ can be expanded as power series in the bulk velocity
\begin{equation}
f_i^{0} = A + Bu_{\alpha}e_{i \alpha} + Cu^2 +Du_{\alpha} u_{\beta}e_{i
\alpha} e_{i \beta}
 + G_{\alpha\beta}e_{i\alpha} e_{i \beta},
\end{equation}
\begin{equation}
f_0^{0} = A_0 + C_0u^2,
\end{equation}
\begin{equation}
g_i^{0} = H + Ku_{\alpha}e_{i \alpha} + Ju^2 +
Qu_{\alpha}u_{\beta}e_{i\alpha} e_{i \beta},
\end{equation}
\begin{equation}
g_0^{0} = H_0 + J_0u^2.
\end{equation}
The expansion coefficients $A,B \ldots$ are determined by
\begin{equation}
\sum_i f_i^{0} = n
\label{eqn10}
\end{equation}
\begin{equation}
\sum_i f_i^{0}e_{i \alpha}= n u_{\alpha},
\label{eqn11}
\end{equation}
\begin{equation}
\sum_i f_i^{0} e_{i\alpha}e_{i\beta} =
P_{\alpha \beta} + n u_\alpha u_\beta,
\label{eqn12}
\end{equation}
\begin{equation}
\sum_i g_i^{0} = \varphi,
\label{eqn13}
\end{equation}
\begin{equation}
\sum_i g_i^{0} e_{i\alpha}= \varphi u_\alpha,
\label{eqn14}
\end{equation}
\begin{equation}
\sum_i g_i^{0} e_{i\alpha}e_{i\beta}
= \Gamma \Delta \mu \delta_{\alpha \beta} + \varphi u_\alpha u_\beta.
\label{eqn15}
\end{equation}
where $P_{\alpha \beta}$ and $\Delta \mu$ are given by (\ref{eqn5}) and
(\ref{eqn3}) respectively and $\Gamma$ is a mobility. Note that the
conditions (\ref{eqn10}), (\ref{eqn11}), and (\ref{eqn13})
correspond to local conservation of density, momentum and density
difference respectively. Explicit expressions for the expansion
coefficients are given in Appendix A.

Expanding (\ref{eqn6}) and (\ref{eqn7}) to order $(\Delta t )^2$ 
and using (\ref{eqn10})--(\ref{eqn15}) leads
to the macroscopic equations
\begin{equation}
\partial_t n +\partial_\alpha( n u_\alpha) = 0 ,
\label{eqn20}
\end{equation}
\begin{equation}
\partial_t (n u_\beta) + \partial_\alpha( n u_\alpha u_\beta )=
-\partial_\alpha P_{\alpha \beta}
+\nu \nabla^2(n u_\beta)
+ \partial_\beta \{ \lambda (n) \partial_{\alpha}(n u_{\alpha})\},
\label{eqn21}
\end{equation}
\begin{equation}
\label{eqn22}
\partial_t \varphi + \partial_\alpha (\varphi u_\alpha )=
\Gamma \theta \nabla^2 \Delta \mu - \theta \partial_\alpha \left(
\frac{\varphi}{n} \partial_\beta P_{\alpha \beta} \right),
\end{equation}
where
\begin{eqnarray}
\label{eqn23}
\theta=(\Delta t) c^2 \left(\tau_2 - 1/2\right),&
\nu=\frac{(2\tau_1 -1)}{6} (\Delta t)c^2,
&\lambda(n)=(\tau_1 -\frac{1}{2})\Delta t(\frac{c^2}{2}-\frac{d
p_0}{dn}). \nonumber \\
&&
\end{eqnarray}

It is apparent from these equations of motion that the fluid will
evolve through two competing growth mechanisms. The first of these is
diffusion, described by eqn.~(\ref{eqn22}) with $\vec{u}=0$. The second
is bulk momentum transfer or hydrodynamic flow described by the
Navier-Stokes equation~(\ref{eqn21}). The relative importance of
diffusive and hydrodynamic flow can be altered by varying
the viscosity $\nu$ through changes in $\tau_1$ (see equation~(\ref{eqn23})).
For high viscosities the velocities remain sufficiently small that
hydrodynamic flow is irrelevant and the evolution of the
microstructure is diffusive. For low viscosities,
however, the Reynolds number becomes larger and
hydrodynamic effects can dominate in changing the domain
morphology. Simulations run from the same initial conditions but with high
or low viscosities enable us to build up a very clear picture of the
effects of hydrodynamics on the domain growth.

\section{50:50 composition: metastability and the effect of
hydrodynamics}

We first describe the behaviour of a symmetric binary fluid with a 
ratio of number densities $n_A:n_B$ of
$50:50$ when it is quenched from a disordered state into the ordered
region. Our aim is to compare the path of the spinodal decomposition
for different values of the surface tension and of the
viscosity. Initial results for 
this concentration have appeared elsewhere\cite{OG97}.

For all runs the system was initialised with $n=1.0$ and
$\varphi$ chosen randomly between --0.5 and 0.5. It was then quenched to a
final state defined by parameters
$a=-1$, $b=\zeta=1$. The simulations were run with $\Delta t=0.004$,
$c=1$, $\tau_2= 0.7786751$ and $\Gamma=1$. 
The system size was $128\times 128$ and
simulations were typically run for $10^5$ time steps. This took
approximately $10$ days  on a DEC Alpha Workstation.
Smaller lattice sizes gave similar results.

For $\kappa$ sufficiently negative the equilibrium state is a lamellar
phase. After a quench at high viscosity 
the fluid gets stuck in a
metastable state of lamellae which have approximately the
correct width, but which are tangled. For low viscosity, however,
hydrodynamic flow builds up and 
can remove topological defects
from the system and allow the fluid to reach equilibrium.

\begin{figure}
\inseps{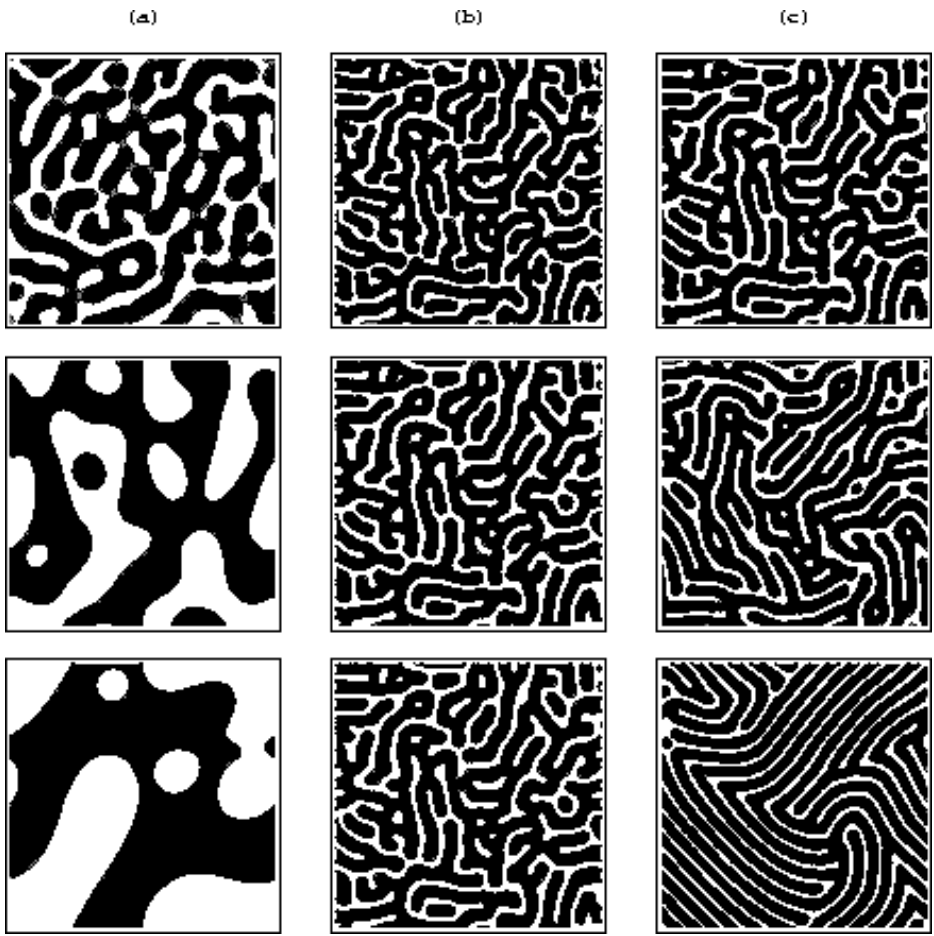}{0.8}
\vskip 10mm
\caption{Snapshots of the growth of domains with time for a binary
mixture symmetric in composition. 
Grey-scaling from black $\Rightarrow$ white corresponds to $\varphi=-1
\Rightarrow \varphi=1$.
Each column
represents a different physical situation: (a)
a quench to the homogeneous two-phase region $\kappa > \kappa_c$;
(b) a quench to the
lamellar phase $\kappa < \kappa_c$ in a high viscosity fluid. The
lamellae form in a tangled pattern which becomes frozen in time;
(c) a quench to the same value of $\kappa$
as (b) but for a low viscosity fluid.
Hydrodynamic modes allow the lamellae to reorder giving, locally,
well-defined striped regions.}
\label{figure1}
\end{figure}

Evidence for these conclusions is shown in
Figure~\ref{figure1} where snapshots of the domain growth are 
compared for three
different sets of parameters: 
\begin{itemize}
\item[(a)] high viscosity, two-phase coexistence ($\tau_1=50$, $\kappa=0.1$),
\item[(b)] high viscosity, lamellar equilibrium ($\tau_1=50$, $\kappa=-0.85$),
\item[(c)] low viscosity, lamellar equilibrium ($\tau_1=0.585$, $\kappa=-0.85$).
\end{itemize}

Consider first (a) which is at a value of $\kappa$ that corresponds to
bulk phase separation and a value of $\tau_1$ that suppresses
hydrodynamic flow. After initial transients sharp domains form. These
are more elongated than for a system with $\zeta=0$ but they grow
continuously and become more isotropic. An exponent $\alpha$
characterising the growth is commonly defined by
\begin{equation}
R(t) \sim t^{\alpha}
\end{equation}
where $R(t)$ is a length scale, which we take as the inverse of the
first moment of the circularly averaged structure factor, and $t$ is
the time\cite{B94}. At late times $\alpha$ converges to 
$1/3$ as expected for Lifshitz-Slyozov diffusive
growth in a binary system. Runs at different values of $\kappa$
indicate that as $\kappa$ is decreased an increasingly long time is
taken to reach the $1/3$ regime\cite{OG97}. 

For $\kappa < \kappa_c \sim -0.8$ there is a qualitative change in the
behaviour of the system as shown in
Figure~\ref{figure1}b. These results were obtained for the same
value of $\tau_1$ but with $\kappa=-0.85$ where the equilibrium is the
lamellar state. Now the system forms
portions of lamellae of width approximately equal to the equilibrium
value. These lamellae are tangled in a way that depends on the initial
conditions. The diffusive growth to the tangled phase 
is slow, possibly logarithmic,
and once the glassy phase has been reached there is no further
discernable movement on the time-scale of the simulation.
For smaller $\kappa$ the lamellae are thinner and the system freezes
sooner.  

Remaining with the value $\kappa=-0.85$ we then ran the same
simulation (starting from the same initial conditions) for
$\tau_1=0.585$. The lower value of $\tau$ implies a lower viscosity
and the possibility of hydrodynamic flow at earlier times.
The initial behaviour was the same as for case (b)
with a glassy lamellar phase being formed. However at a later time
$t~$ there was a significant change in the
domain pattern. This is most easily seen from the snapshots in 
Figure~\ref{figure1}c where it is apparent that the 
topological frustration of the glassy
phase is being removed and the lamellae are lining up in the global
equilibrium state.

An example of the effect of hydrodynamic flow on the defects is shown
in Figure 2. 
The lamellae are initially too wide. Hydrodynamic forces tend to
lengthen them causing the broken lamellar to join and the stripes to
buckle. At high viscosities the defect does not disappear.

\section{60:40 composition: separation to coexisting lamellar and bulk phases}

We next consider a concentration ratio $n_A:n_B=60:40$. Now there is too
little of the $B$-phase 
to form lamellae of the correct width throughout the system. 
We shall show that the
equilibrium state corresponds to phase separation into a
lamellar region coexisting with a bulk $A$-phase. As before equilibrium
can only be reached with the help of hydrodynamic flow.

Simulations were again run at different values of $\tau_1$ and
$\kappa$ to provide a comparison. Snapshots of the time evolution are
shown in Figure~\ref{figure3} for the same parameters as those used
for the symmetric mixture considered in Section IV.
Figure 4 is a double logarithmic plot of the variation of the domain
size with time comparing the growth in the three different cases.

Column (a) of Figure 3 shows the path to bulk phase separation 
for $\kappa=0.1$.
Droplets form by spinodal decomposition and then grow by
Lifshitz-Slyozov diffusion. The curvature term in the free energy
becomes less important for larger droplets and they become more
circular. The data in Figure 4 is consistent with the expected growth
exponent $\alpha=1/3$ at late times.

Figure 3b compares a simulation for $\kappa=-0.85$, a value for which
the lamellar phase is stable in the symmetric mixture. The value of
$\tau_1$ is chosen so that only diffusive growth is possible. It is 
apparent from
Figure \ref{figure3}b that the final state is a mixture of droplets and short
lamellae. There are no further discernable changes in morphology on the
time scale of the simulation as confirmed in Figure 4. 

Evidence that
this droplet state is metastable is provided when the same simulation
is run with a low viscosity. 
Bulk fluid flow now allows the droplets to join to form lamellae and 
align. A surprising amount of movement is seen leading to the growth
process shown in Figure~\ref{figure3}c. 
In Figure 4 the onset of hydrodynamic flow in this system 
is marked by a rather sharp
increase in the measured length scale.

The final state is 
coexistence between a lamellar and a bulk A-phase. 
The lamellae order in a distinctive spiral around the hole of A-phase,
to minimise the elastic free energy which would be generated by
incorrect lamellar spacings. 

Finally we remark that very 
similar results were obtained for a concentration ratio of 
$n_A:n_B=70:30$.

\section{90:10 composition: the droplet phase}

Finally we describe phase separation in a highly asymmetric binary
fluid with $n_A:n_B = 90:10$. Now the equilibrium for $\kappa$
sufficiently negative comprises droplets of $B$ in $A$. 
Figure 5 compares results for two different values of $\kappa$:
\begin{itemize}
\item[(a)] high viscosity, two-phase coexistence ($\tau_1=50$, $\kappa=0.1$)
\item[(b)] high viscosity, lamellar equilibrium ($\tau_1=50$, $\kappa=-0.85$)
\end{itemize}
The main difference between the growth processes in Figures 5a and 5b
is a direct consequence of the final equilibrium state. For
$\kappa=0.1$, once the domains have formed they continue to grow
slowly by the diffusion of material between them. This is the
Lifshitz-Slyozov growth process, driven by the difference in chemical
potential between droplets of different size. For $\kappa=-0.85$ however
there is a preferred size for droplets, set by the competition between
the surface tension and the curvature energy, and growth stops once the
droplets have reached this size. A second difference is that for the
negative value of $\kappa$ droplets form much more quickly in the
early stages of growth. This is a consequence of the reduced surface tension.

Runs for a low viscosity and negative $\kappa$ showed a negligible 
influence of hydrodynamics on growth for this concentration.
This is as expected. Hydrodynamic flow can act to make a droplet
circular because of the pressure difference between points of
different curvature. However once the drops are circular 
hydrodynamic flow cannot directly lead to droplet coalescence although
it may speed up the diffusive growth\cite{T95}.

\section{Conclusions}

We have shown that there is a wealth of interesting behaviour as even
rather simple structured fluids attain equilibrium. Competition
between diffusive and hydrodynamic modes lead to final states
dependent on the dynamic parameters of the system. In particular we
have emphasised that hydrodynamic flow is often important in allowing
a system to reach its equilibrium state. 

The results were obtained using lattice Boltzmann simulations. The
approach has two particular advantages in this context. Firstly
equilibrium is determined by a free energy which is an intrinsic part
of the simulation so the structure of the equilibrium state can be chosen
rather naturally. Secondly the viscosity of the fluid can be tuned
over a wide range. We caution however that the lattice Boltzmann
evolution is not described by an H-theorem and thus no proof exisits
that the path to equilibrium is a physical one.

Lattice Boltzmann simulations do not intrinsically include noise
although this can be added in an ad hoc way. An important question is
whether such fluctuations can relax the disordered lamellar states. We
ran simulations aimed at investigating this but found no
effect of noise on the glassy structure. In real two-dimensional
smectics fluctuations destroy the lamellar order. However, for this
model we expect a stable lamellar phase, both because of the lack of
fluctuations and because we impose a mean-field free energy.

Our conclusions are in broad agreement with those of Bahiana and
Oono\cite{BO90} who considered the spinodal decomposition of a model
of block copolymers designed to give a lamellar equilibrium. Despite
the long-range interactions in the model a tangled lamellar phase was
formed after a quench. This could be ordered by including terms
approximating hydrodynamic flow in the simulation. Other related
work\cite{PD95} uses a Langevin approach which includes hydrodynamics
to model phase separation in microemulsions. The domain growth slows
as the surfactant density is increased. This method provides an
alternative numerical way to study complex fluids and it would be
interesting to gain a fuller understanding of the applicabilities of
the different approaches.

There are many questions that remain to be considered. These include
the effect of confinement or shear on the phase separation, dynamical
asymmetry in the viscosities of the two fluid components, and the
r\^ole of a very low diffusivity, which has been shown to alter the
path to bulk phase separation in binary fluids\cite{T94}. 
Work is in progress to
include a surfactant as a third phase rather than modelling its effect
by changing the surface tension. Extensions to three dimensions are
highly desirable. These are feasible but at the limit of current
numerical resources.\\
~\\
\noindent{Acknowledgements}

We should like to thank P.V. Coveney and A. Wagner for
helpful comments. JY acknowledges support from the EPSRC  (grant no. GR/K97783)
and NATO (grant no. CRG950356).

\section*{Appendix A}
\label{5}

A suitable choice of the coefficients in the expansions of the lattice
Boltzmann equilibrium distributions (10)--(13)
consistent with the conditions (14)--(19) is
\begin{eqnarray}
&A  = \frac{5}{4}\left ( \frac{a}{2}\varphi^2 + \frac{3b}{4}\varphi^4
-\kappa \varphi\left( \nabla^2 \varphi\right ) + 
\xi\varphi\left(\nabla^2\right)^2\varphi
+ \frac{\zeta}{2}\left(\nabla^2\varphi\right)^2\right)/ (12 c^2),
& \nonumber \\
&A_0 = n - 16A, \qquad
B = 5n/(12c^2), \qquad
& \nonumber \\
&C_0 = -2n/(3c^2), \qquad
C = - 5n/(24c^2), \qquad
D = 5n/(8c^4), \qquad
& \nonumber \\
&G_{xx} = -G_{yy} =  
\frac{5\kappa}{2}\left ( \left(\partial_x\varphi\right)^2 - 
\left(\partial_y\varphi\right)^2\right)/8c^4 +
5\zeta\left ( \partial_y\varphi\partial_y\left(\nabla^2\varphi\right)
-\partial_x\varphi\partial_x\left(\nabla^2\varphi\right)\right)/8c^4,
& \nonumber \\
&G_{xy} = 5\kappa\partial_x\varphi\partial_y\varphi - 
5\zeta \left\{ \partial_x\varphi\partial_y\left(\nabla^2\varphi\right) +
\partial_y\varphi\partial_x\left(\nabla^2\varphi\right)\right \},
& \nonumber \\
&H_0 = \varphi - 4H, \qquad
H=5 \Gamma \Delta \mu/(12 c^2), \qquad
K =  5 \varphi/(12c^2), \qquad
& \nonumber \\
&J_0 = -2 \varphi/(3c^2), \qquad
J = -5 \varphi/(24c^2), \qquad
Q = 5 \varphi/(8c^2). \qquad
& 
\label{106}
\end{eqnarray}

\begin{figure}
\vskip +5mm
\inseps{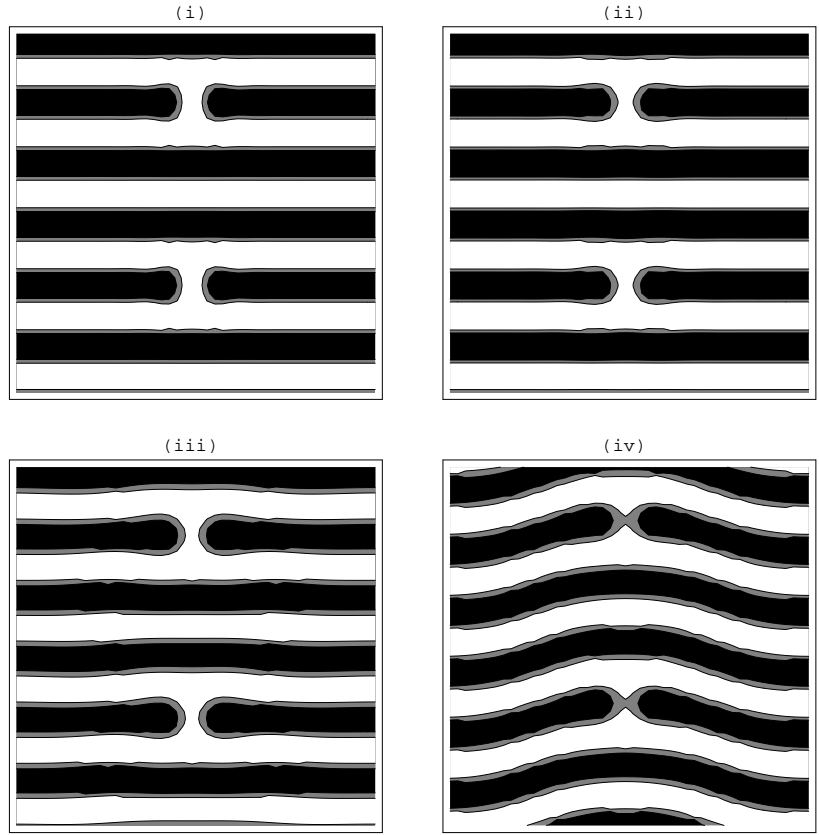}{0.7}
\vskip -5mm
\caption{Defect being removed by flow. The lamellae are initially 
too wide. Hydrodynamic forces tend to
lengthen them causing the broken lamellar to join and the stripes to
buckle. Grey-scaling from black $\Rightarrow$ white corresponds to $\varphi=-1
\Rightarrow \varphi=1$.
}
\label{figure2}
\end{figure}

\begin{figure}
\inseps{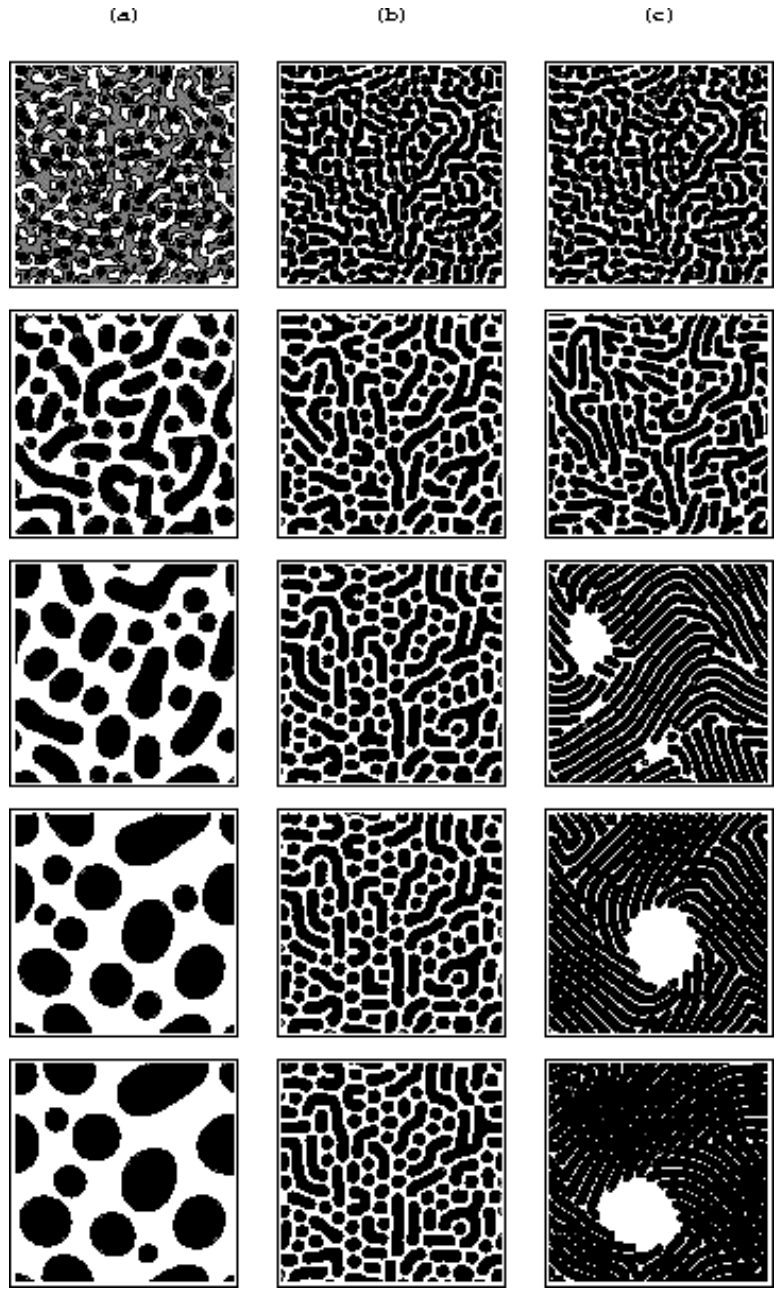}{0.8}
\vskip 10 mm
\caption{Snapshots of the growth of domains with time for a binary
mixture slightly asymmetric in composition.
Grey-scaling from black $\Rightarrow$ white corresponds to $\varphi=-1
\Rightarrow \varphi=1$.
(a) A quench to the homogeneous two-phase region $\kappa > \kappa_c$;
(b) a quench to the
lamellar phase $\kappa < \kappa_c$ in a high viscosity fluid. Short 
lamellae form in a pattern which becomes frozen on the time scale of
the simulation;
(c) a quench to the same value of $\kappa$
as (b) but for a low viscosity fluid.
Hydrodynamic flow allows the system to attain its equilibrium of 
coexisting lamellar and bulk-$A$ phases.}
\label{figure3}
\end{figure}

\begin{figure}
\vskip -200mm
\inseps{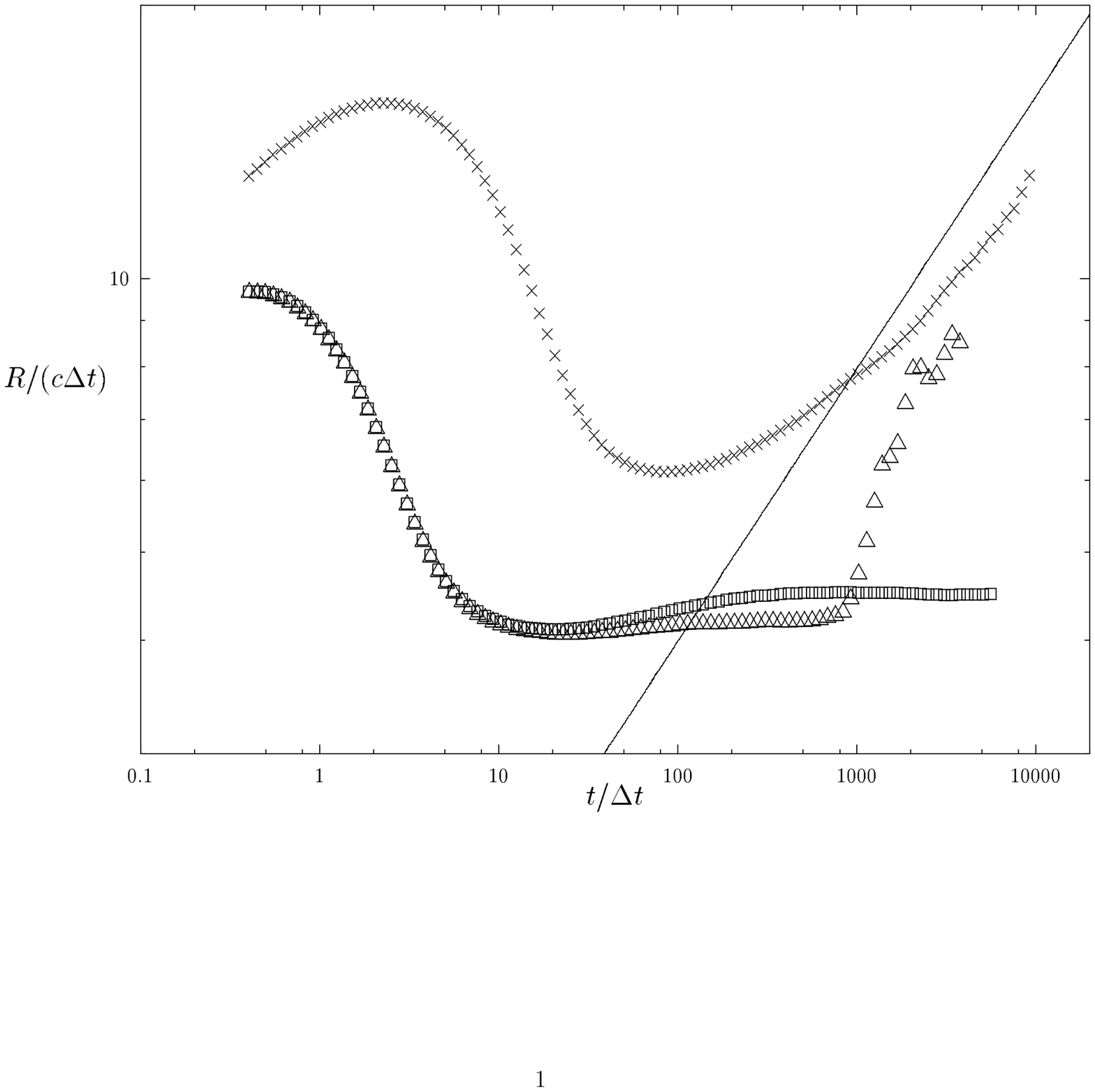}{0.7}
\vskip -50mm
\caption{Double logarithmic plot of the evolution 
of the inverse first moment of the
structure factor as a function of time 
for each of the simulations shown in Figure 3: 
($\times$) bulk phase separation,
($\Box$) a quench to the lamellar phase with high viscosity, 
($\triangle$) 
a quench to the lamellar phase with low viscosity. 
The straight line has slope $1/3$.}
\label{figure4}
\end{figure}

\begin{figure}
\inseps{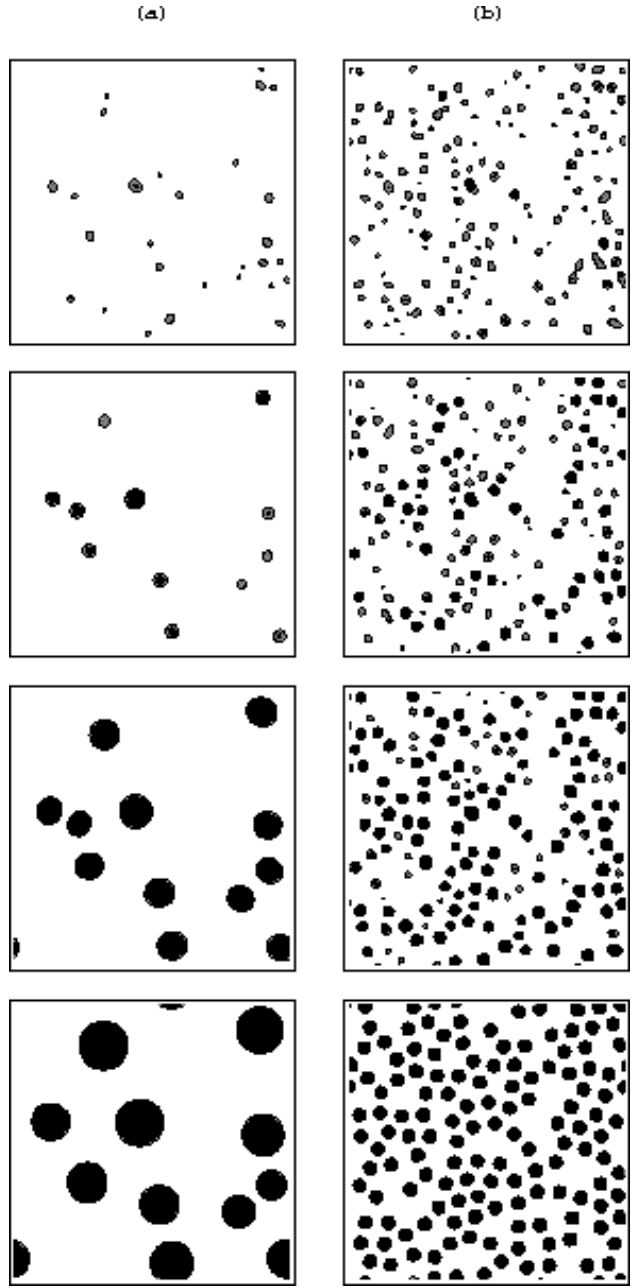}{0.8}
\vskip +10mm
\caption{Snapshots of the growth of domains with time for a binary
mixture highly asymmetric in composition. Grey-scaling from 
black $\Rightarrow$ white corresponds to $\varphi=-1
\Rightarrow \varphi=1$.
(a) A quench to the homogeneous two-phase region $\kappa > \kappa_c$;
(b) a quench to the
droplet phase $\kappa < \kappa_c$ where the final length scale is set by the
competition between the surface tension and the curvature energy.} 
\label{figure5a}
\end{figure}

\end{document}